\documentclass{aa}
\usepackage{epsfig}
\newcommand{\kms}{~km~s$^{-1}$}
\newcommand{\Mdot}{\mbox{$\dot M$}}                 
\newcommand{\Msolyr}{~M$_{\odot}$\,yr$^{-1~}$}      
\begin{document}
\title{The envelope of IRC+10216 reflecting the galactic light 
       \thanks{Based on observations made with the Antu 8-m VLT (ESO; 
        program 63I-0177A), with the Canada France Hawaii 3.6-m Telescope
        (CNRS, NRC, UH)
        and the 1.20-m telescope of Haute-Provence Observatory (CNRS)}
}
\subtitle{$UBV$ surface brightness photometry and interpretation}
\author{N.~Mauron\inst{1} , P. de Laverny\inst{2}, B. Lopez\inst{2}}
\offprints{N.~Mauron}

\institute{Groupe d'Astrophysique, UMR~5024 CNRS, Case CC72, 
Place Bataillon, 34095 Montpellier, France\\ 
 \email{mauron@graal.univ-montp2.fr}
      \and 
 Observatoire de la C\^ote d'Azur, D\'epartement Fresnel, UMR~6528 CNRS, BP 4229, 
06304 Nice, France\\             
\email{laverny@obs-nice.fr, lopez@obs-nice.fr}
     }

\date{Received 8 July 2002 / Accepted 27 December 2002}

\abstract{
We present and analyse new optical images of the dust envelope surrounding the
 high mass-loss carbon star IRC+10216. This envelope is seen due to
external illumination by  galactic light. Intensity profiles and
colors of the nebula were obtained in the $UBV$ bandpasses. The data
are compared with the results of a  radiative transfer model calculating
  multiple scattering of  interstellar field photons by  dust grains 
  with a single radius. 
The data show that the observed radial shape of the nebula, especially its 
half maximum radius, does not
depend on wavelength (within  experimental errors), suggesting that
grains scatter in the grey regime, and this is further supported by the
plateau colors being close to those of the ISRF  as given by Mattila 
(\cite{mattila80a}).
A grain radius of 0.16 $\mu$m with  envelope parameters as
proposed by Groenewegen (\cite{groenewegen97}) can reproduce this 
achromatism of shape  and color  characteristics.  However, there 
remain substantial discrepancies between model and
observations concerning the absolute intensity of the nebula 
and its radial shape. 
Some of these discrepancies disappear if one adopts a small grain size 
($\sim$ 0.05\,$\mu$m), or if one assumes a lower dust mass loss rate for the
outer layers ($\theta \ge 20$$''$, corresponding to 1000 years ago). 
Within the framework of our simple model, we cannot determine a ``dominant''
grain size. Future more sophisticated models will have to take into account 
grain size distribution, and
also explore complicated issues like the effects of grain porosity and/or
asphericity on scattering, the influence of the envelope small-scale structure
on the radiative transfer, and the possibility of a field anisotropy.
For the same reasons, it is not presently feasible to
establish with confidence whether the interstellar radiation field in the 
visible is  significantly different in strength at the location of 
IRC+10216 compared to the usually adopted one in the solar neighbourhood
}
\titlerunning{The envelope of IRC+10216}
\authorrunning{ N. Mauron, P. de Laverny, B. Lopez}
\maketitle

\section{Introduction}

Stars of intermediate initial  mass  can have
very high mass-loss rates when evolving on the asymptotic giant 
branch (AGB). 
This mass loss produces extended, dusty, optically
thick envelopes that strongly mask the central star in the
visible domain, possibly leaving no optical counterpart.
 However, these envelopes are externally illuminated by the stars of 
the galactic disk. If deep enough, a direct image of the envelope can 
show the reflection of this external photon source
 by  circumstellar grains. Two such cases of AGB reflection nebulae 
are presently known. The best case is  the nearby 
 carbon star IRC+10216, with an exceptionally massive circumstellar
  envelope that is clearly seen
 in the ambient galactic light in  the  $B$ and $V$ bands (Mauron \& 
 Huggins \cite{mauron99}). The envelope is detected up to 
 about 3$'$ from center,
  and appears fairly round, which is qualitatively consistent with 
a  spherically symmetric dust shell and an
isotropic galactic field (at least in $B$ and $V$). 
There is also a central {\it plateau} of diameter $\sim$ 20$''$,
 (i.e. no central peak), which is prominent in  $B$ and
proves that the source of photons is not the central star.
The second case is CRL 3116,  another carbon AGB object at about
1 kpc from the Sun (Crabtree \& Rogers \cite{crabtree92}), but 
this object is much less studied than IRC+10216.\\

In the present paper, we follow the works of Martin \& Rogers 
(\cite{martin87}, hereafter MR87) and  Crabtree \& Rogers (\cite{crabtree92}),
 and our goal is to make a quantitative investigation
of these reflection nebulae in order to derive information 
on  the interstellar radiation field (ISRF) and/or 
the circumstellar dust properties.
Briefly, the ISRF is due to the integrated light of galactic disk stars 
 and  a diffuse galactic light component. It is an important quantity
involved  in many problems concerning the 
interstellar medium or circumstellar shells, such as
 grain heating, their far-infrared emission and 
photochemistry (e.g., Mathis, Mezger \& Panagia \cite{mathis83},   
van Dishoek \cite{vandishoek94}, Glassgold \cite{glassgold96}). The 
intensity of the local ISRF (measured from Earth) is relatively 
well known in the  visible range,  but  the situation is more controversial 
in the  ultraviolet, with differences between  authors of at least 
a factor of 2  at 1500\AA~ (see Fig. 2 of van Dishoek \cite{vandishoek94}).
Concerning the spatial ISRF variations 
within the solar neighbourhood, one  expects rather small variations,
except for peaks  close to very  luminous stars,
especially in the far ultraviolet region (Habing \cite{habing68}, 
Jura \cite{jura74}).\\

 Martin \& Rogers (\cite{martin87})  found that in order to account for the 
$V$ surface brightness of the IRC+10216 plateau, which was
relatively faint compared to their model, a possible solution
 was to assume an ISRF smaller than that of Mathis et al. 
 (\cite{mathis83}) 
by a factor of  $\sim$ 2.5. 
 They suggested that a faint ISRF could be due 
to the height of this object above the galactic plane, which is
$z = 90$\,pc for an assumed distance  $d = 130$\,pc. However, 
according to Mattila (\cite{mattila80a}),
the dependance on $z$ is predicted to be much smaller than a factor of 2,
at least in the optical domain. Another 
indication that the ISRF may have no strong  minima
comes from the COBE observations of interstellar dust at high
galactic latitude: the narrow range
of dust temperature suggests ISRF variations smaller 
than 30\% (Lagache et al. \cite{lagache98}).\\

 Martin \& Rogers  also noted two other possible  explanations for
the low  plateau  $V$-band brightness: the first is to assume
a grain albedo lower than originally adopted in their model;
the second is to envisage an even smaller 
grain radius than assumed  in their model (0.05 $\mu$m) and to 
increase the mass loss rate in the
external layers. However more recent studies of the dust of
IRC+10216 point to significantly larger grain sizes: a
dominant radius of 0.16 $\mu$m is found by Groenewegen 
(\cite{groenewegen97}, GR97
hereafter). Distribution sizes including grains as large as 
0.2 or 0.4 $\mu$m are also favoured by Ivezic \& Elitzur (\cite{ivezic96})
 and by Skinner, Meixner \& Bobrowsky (\cite{skinner98}).\\

These considerations led us to obtain additional images 
of the ISRF-reflecting nebula around IRC+10216. Our goal is 
to reanalyse the intensity, colors, size, and shape of this nebula.
In particular, this paper 
presents new surface brightness measurements  in the 
three $UBV$ bandwidths (a photometric sequence of field stars 
is given in appendix A).
 Details about the observations and the data 
reduction  are given  in Sect.~2, and the results are presented in
Sect.~3. A specific  Monte-Carlo 
radiative  transfer code was  developped for this program and 
is described in Sect.~4, with complementary considerations in
Appendix B. The data are quantitatively interpreted and discussed 
in Sect.~5, and a summary of this work is presented in the 
conclusion.\\ 

\section{Observations and reductions}

The first set of observations was carried out at Haute-Provence
 Observatory with the 1.20-m telescope,
 equipped at the Newtonian focus  with a CCD  camera for $UBVRI$ imaging.
The chip is a thin back-illuminated  1024$^2$ Tektronix, with
a read-out noise of 8.5~$e^{-}$  and a gain of 3.5 $e^{-}$~ADU$^{-1}$.  
This instrument provides a field of 11.7$'$$\times$11.7$'$ with a 
scale of 0.6845 arcsec pixel$^{-1}$. More details  
 can be found in Ilovaisky (\cite{ilovaisky97}).
The nights were clear but the seeing was 2$''$. 
 Flat-fields were done on the dome and on the sky. 
The $UBV$ images of IRC+10216  obtained are listed in Table~\ref{table1logobs}.
Several images were also taken each night 
on the calibration fields NGC 2403 (Zickgraf et al. \cite{zickgraf90})
and M13 (Arp \& Johnson \cite{arp55}).
All CCD frames were reduced in the standard way, with bias
substraction and division by  averaged 
flat-fields. These data permitted the establishment of a
$UBV$ photometric sequence containing 16 stars in the field of IRC+10216
(details in Appendix A).\\

The CFHT observations are listed in Table~\ref{table1logobs} and were described by Mauron 
\& Huggins (\cite{mauron99}). They resulted into two deep $B$ and $V$ images,
with a field of 8.7$'$$\times$8.7$'$  and a scale of 0.436 arcsec 
pixel$^{-1}$.  Because
no photometric standards were observed at CFHT, calibration  was 
achieved by considering unsaturated stars of the OHP sequence.\\

The VLT observations were done in service mode with the Antu 8-m telescope,
equipped with the FORS1 focal reducer (Table~\ref{table1logobs}). The detector is
a 2048$^2$ thinned 24\,$\mu$m pixel Tektronix chip.
The field of view is 6.83$'$$\times$6.83$'$  with a pixel 
size of 0.200 arcsec pixel$^{-1}$. More details on the instrumentation
can be found in the FORS1 Manual at ESO Web site.
The usable data consists of three 8-min $U$ 
frames and two 2-min $V$ frames. 
A number of other frames in $UBV$ 
were acquired on different nights, but had to be rejected 
for this study due to an excessively bright sky or  poor
photometric  conditions. The  frames were merged into one $U$ and 
one $V$  flat-fielded image [This 4 mn VLT $V$-band image is much less
deep  than the CFHT one presented in Mauron \& Huggins (\cite{mauron99})
and is not shown here; for the $U$-band image, see below]. 
 During the nights of
observation, one or two exposures per filter were supplied of
calibration fields (Mark\,A, SA\,101 and PG\,0942-029, 
detailed on the FORS ESO Web site). These exposures did not allow 
an independant determination of the atmospheric extinction  coefficients 
of these nights, but if plausible values are assumed 
($k_U$=0.62, $k_V$=0.14 mag/airmass),
one finds that the $U$ and $V$  photometry
of the field stars  agree very well with the OHP results, with
differences of  0.04 mag in $V$, and 0.03 mag in $U-V$. Therefore,
all measurements on the nebula were calibrated using
the OHP $UBV$ sequence.\\

\begin{table}
   \caption[]{Log. of observations of IRC+10216}
   \label{table1logobs}
   \begin{flushleft}
   \begin{tabular}{lrrrr}
   \hline
   \hline
\noalign{\smallskip}
    Telescope& Date~~~~ & Band & Exposures\\
\noalign{\smallskip}
     \hline
\noalign{\smallskip}
 OHP 1.2m  & 27-28 Jan 1998   &  $B$      & 4 x 1 hr  \\
 OHP 1.2m  & 27-28 Jan 1998   &  $V$      & 2 x 1 hr  \\
 OHP 1.2m  & 28-29 Jan 1998   &  $U$      & 5 x 1 hr  \\
 CFH 3.6m  & 17-18 Feb 1998   &  $V$      & 4 x 20 mn \\
 CFH 3.6m  & 17-18 Feb 1998   &  $B$      & 2 x 20 mn \\
 VLT 8.0m  & 6 Jun \& 9 May 1999   &  $U$      & 3 x 8 mn\\
 VLT 8.0m  & 9  May 1999           &  $V$      & 2 x 2 mn  \\
\noalign{\smallskip}
    \hline 
\noalign{\smallskip}
   \end{tabular}\\
   \end{flushleft}
\end{table}

After bias subtraction and flat-field correction, our goal was
to obtain surface photometry of the  plateau  in the $UBV$ bands,
and to derive the surface brightness profiles. The main difficulty 
is to estimate the background sky  level to be subtracted,
for several reasons. First, the  plateau  is intrinsically faint,
about 4\%  of the sky brightness in $V$. This implies that, if 
an accuracy of $\sim$ 10\% is desired for the  plateau  photometry,
the sky brightness far from the nebula
has to be determined with a relative error of $\sim$ 0.4\%.
The flat field correction should also 
be achieved with this accuracy over large or medium
angular scales, but in practice the corrected images show small but 
non-negligible gradients. These gradients can be corrected in part by
dividing by second order polynomial functions, but residual
variations remain at intermediate scales.
The situation is also complicated by the fact that
there are three bright stars  
in the field: they produce haloes, various ghosts, and 
 significant diffuse parasite light  that is not uniform. 
Nevertheless, these defects depend on the   intrumentation used, 
and we could perform cross-verifications between OHP, CFHT,
and VLT data. The surface photometry of the nebula was obtained with
azimuthally averaged radial profiles. In the $U$-band, the center
position given by the $V$ frames  was adopted . The profiles are  
achieved after masking all stars, 
galaxies, bright stars and their haloes, and 
other obvious defects (see MH99 for more details).

\begin{figure}
\resizebox{\hsize}{!}{\includegraphics{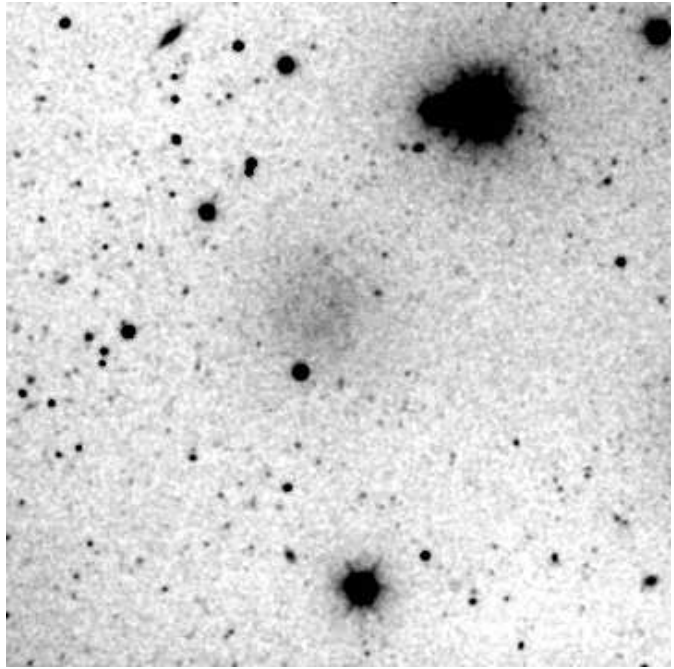}}
\caption{ Composite $U$ image obtained by summing all $U$ frames from
OHP and VLT observations. The field  is 6.0$\times$6.0 arcmin$^{2}$, with
  1.0$''$ pixel. North is up, East to the left.}
\label{fig1uimage}
\end {figure}

\begin{figure}
\centering
\includegraphics[angle=-90,width=8.5cm]{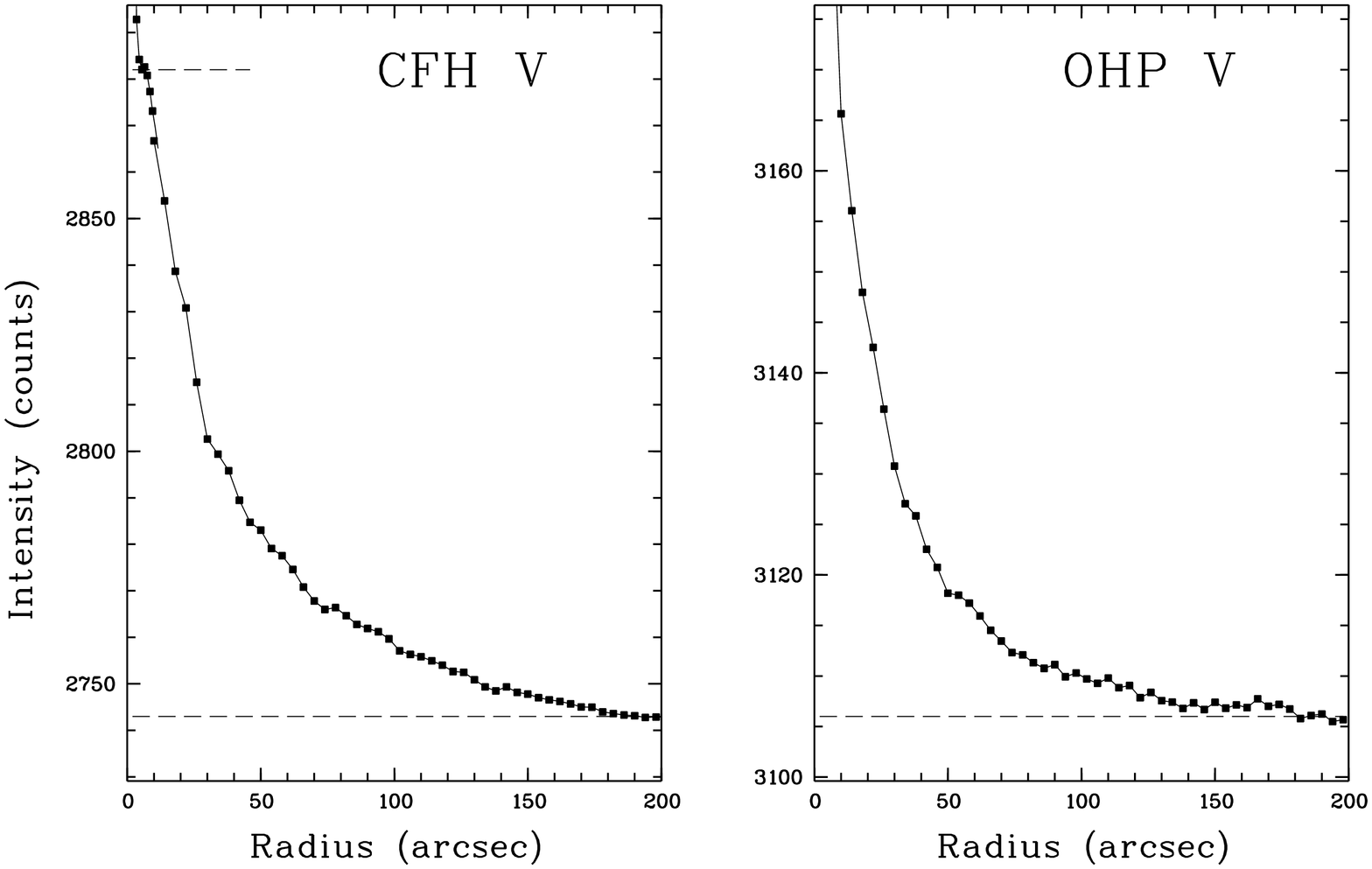}
\includegraphics[angle=-90,width=8.5cm]{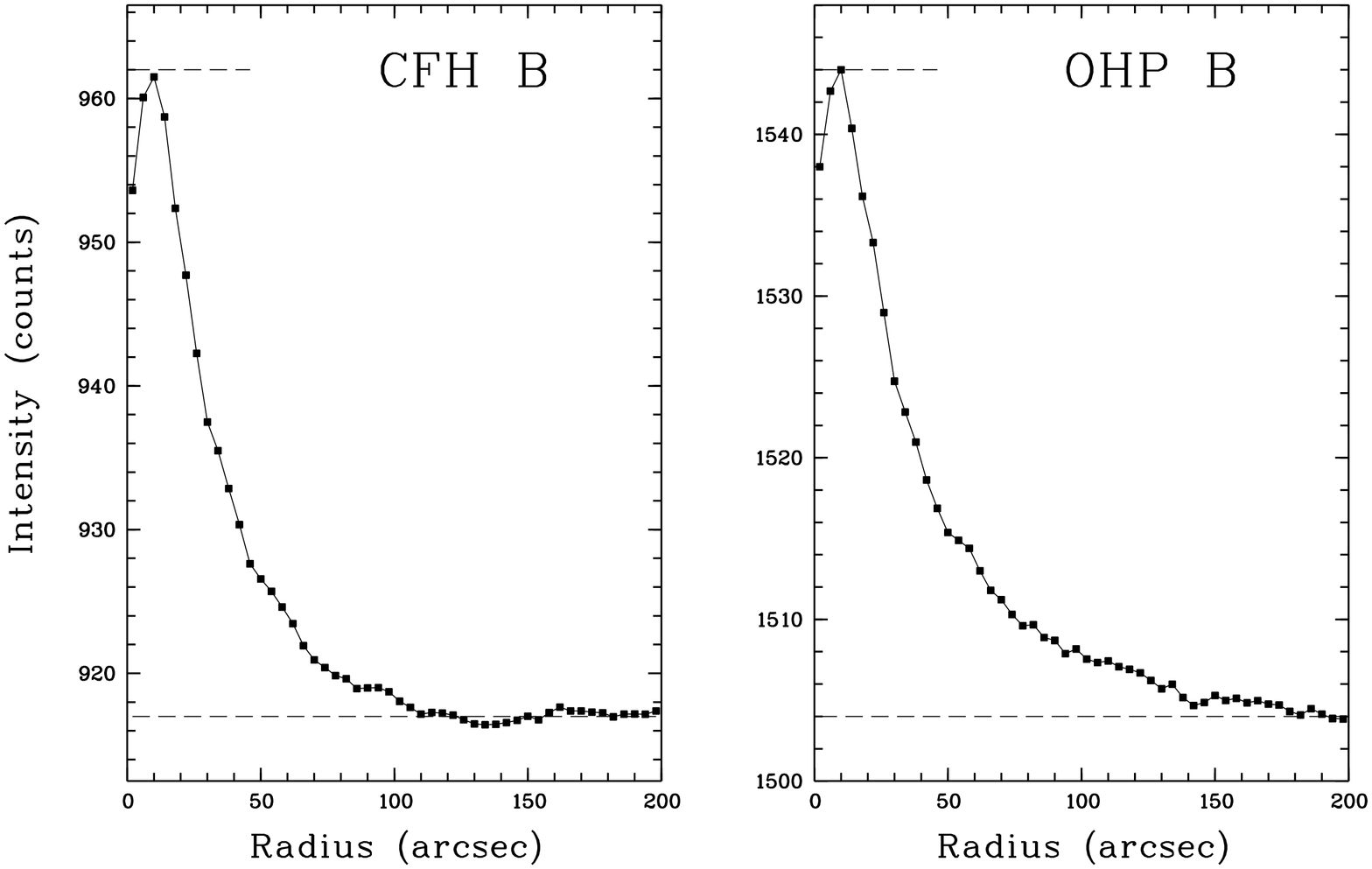}
\includegraphics[angle=-90,width=8.5cm]{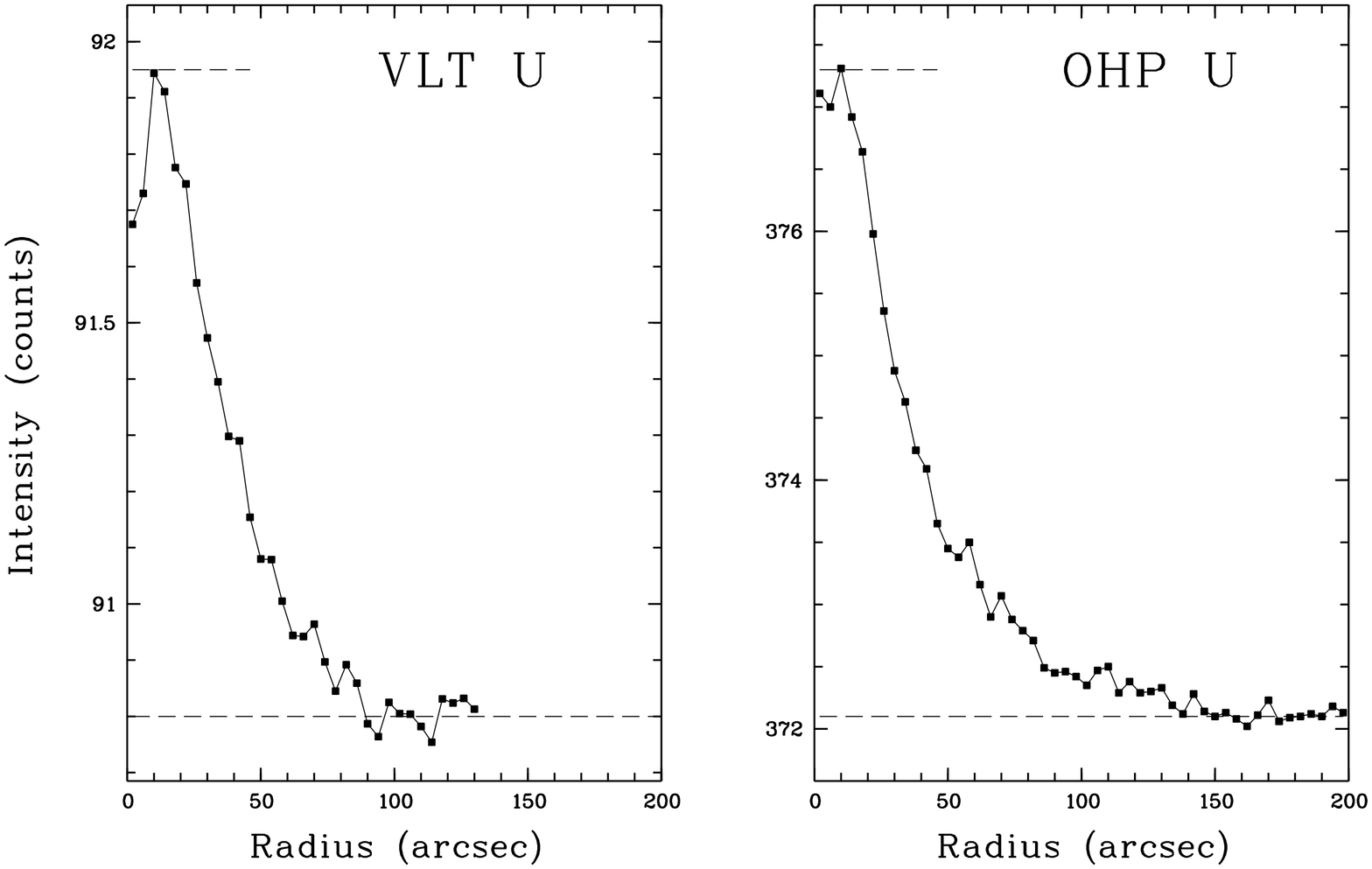}
\caption{ Profiles of the IRC+10216 nebula in $UBV$ bands.
No sky subtraction has been done.  
The ordinate scales have been adjusted so that
the background  level and the nebula ``plateau''
(central zone within a radius of $\sim$ 15$''$)  have in all panels
the same levels,  indicated by dashed horizontal lines}
\label{fig2profils}
\end {figure}

\section{Observational results}

The reduction of all images resulted in the  final
 plateau  $UBV$ surface photometry given in Table~\ref{table2results}. The $UBV$ 
measurements refer to  the top of the  plateau, i.e. at $r = 10$$''$,
on azimutally averaged
profiles. The presence of a central nearly point-like peak due  to radiation
from the star in the CFHT and VLT
$V$ images was taken into account and a small  correction was done 
when estimating the  plateau  maximum intensity at 
$r = 10$$''$. No $V$  plateau  photometry  could be achieved from the OHP data 
because of an excessively  strong  central peak.
 The $B$ and $V$ surface brightnesses of Table~\ref{table2results} are 
 not very different to  previous values given by MH99, and supersede them.
In $V$, the plateau is found to be 0.24 mag brighter than that found by
Crabtree et al. (\cite{crabtree87}).

Figure~\ref{fig1uimage} displays a composite image of the envelope in the $U$ band,
made by summing all the long OHP and short VLT $U$ exposures 
(the summed number of photons per arcsec$^2$ emitted by the nebula 
is comparable in OHP and VLT observations). 
 Comparison with the $B$ and $V$ images displayed in MH99 shows that, in the 
$U$ filter, the nebula may be slightly brighter on the South-west side, 
which is the direction towards the  galactic equator. This 
might be due to the UV-bright OB stars concentrated in the galactic plane, 
but a deeper image is obviously necessary to confirm this asymmetry.

Figure~\ref{fig2profils} shows  azimuthally averaged profiles of the nebula.
Each point corresponds to the signal average over a 4$''$-wide annulus.
No sky subtraction has been done here. The ordinates are CCD counts (ADU)
and the  scales have been adjusted so that both the  plateau  and the sky are
located  at about the same level (dashed line) for any instrument 
and filter.
This scaling could not be done  for  OHP $V$ because the
plateau is not well detected, and this profile is 
displayed here only to show the nebula extension at faint levels.
 This figure shows that
there is no noticeable difference in the nebula shape as a function
of wavelength, given the experimental uncertainties.
It is worth noticing that despite the very low net signals
in the $U$-band, i.e. 1.5 ADU at VLT, 5 ADU at OHP, the $U$ profiles are
well recovered thanks to the averaging over a large number (thousands) 
of pixels located in the annuli. The profiles show that the nebula is 
detected up to $\sim$~100 
or $\sim$~200 arcsec, depending on  instrument and  bandwidth. These
differences are instrumental effects. They reflect both  various depths of the
exposures and  lack of background flatness at very faint levels.

The depth of the central depression is best measured in the $B$ band. In this
band the central core of IRC+10216 is a 
very faint point-like source, with  $B \approx 25.2$ in the CFH frame. 
Its contribution
can be easily subtracted, and the  depression depth
is found to be 17$\pm$3 percent in $B$. Estimating  the depth in $U$ is
much less accurate due to poor S/N in the central bin ($ r < 4$$''$).

The half maximum radius of the profile is also best measured in $B$ and is 
$R_{1/2} = 29$$''$ $\pm$ 2$''$ from OHP and CFH data. 
There is no evidence for a larger $R_{1/2}$ value in $U$, 
taking into account experimental uncertainties (Fig.~\ref{fig2profils}).
Finally in the $V$ band, $R_{1/2}$ is measurable with less accuracy, and
is estimated to be around 27$''$.

\begin{table}
   \caption[]{Results concerning the nebula  of IRC+10216}
   \label{table2results}
   \begin{flushleft}
   \begin{tabular}{ll}
   \hline
   \hline
\noalign{\smallskip}
    Quantity& Result\\
\noalign{\smallskip}
     \hline
\noalign{\smallskip}
$Plateau$ Surf. Bright.  $U$   & 25.85 $\pm$ 0.15   mag. arcsec$^{-2}$\\
$Plateau$ Surf. Bright.  $B$   & 25.70 $\pm$ 0.10   mag. arcsec$^{-2}$ \\
$Plateau$ Surf. Bright.  $V$   & 25.20 $\pm$ 0.10   mag. arcsec$^{-2}$ \\
Color  $U$-$B$                 & 0.15  $\pm$ 0.25               \\
Color  $B$-$V$                  & 0.50  $\pm$ 0.20                \\
Half-light radius in $B$ \& $U$ & 29    $\pm$ 2     arcsec \\
Central depression  in $B$           & 17    $\pm$ 3 percent           \\
\noalign{\smallskip}
    \hline 
\noalign{\smallskip}
   \end{tabular}\\
\end{flushleft}
\end{table}

\section{Modelling}

In order to interpret the observational data, we have to adopt quantitative
characteristics for the  ISRF and the circumstellar dust envelope (CSE),
 and the scattering of light by the dust has to
be calculated. Our approach to the ISRF and the CSE parameters
is to consider the most recent and plausible values derived 
from other studies (although as seen below, significant uncertainties 
exist), and to investigate whether they can explain the observations. 

\subsection{Parameters for the ISRF}

We have considered two different sources for the ISRF. The first
is that of Mathis, Metzger \& Panagia (\cite{mathis83}, MMP83), who
carried out a synthesis of various
measurements and models in order to derive ISRF values  as a function of 
galactocentric distance, from far UV to far IR wavelengths. 
Table~\ref{table3isrf} reproduces the $\lambda$, $f(\lambda)$ values 
of their Table~A3 for the solar neighbourhood (i.e. for $D_G = 10$\,kpc 
as adopted by them). Corresponding apparent $UBV$ magnitudes and 
colors were derived  with the appropriate  formulae of 
Allen (\cite{allen73}), which are indicated in the Table~\ref{table3isrf} notes.

The  second work we considered is that of Mattila (\cite{mattila80a}) who 
calculated synthetic spectra of the integrated stellar light
(ISL) between 3\,000~\AA \, and 10\,000~\AA \, in the solar neighbourhood.  
The diffuse galactic light, another component of the ISRF in addition to 
direct starlight and which comes from light scattered by dust,
is not modelled in that work. However, according to 
Mattila (\cite{mattila80b}),
its effect on the total radiation density averaged over the whole sky
is estimated to be small, of the order of 20\%. Here, we have
chosen to ignore this small component for clarity.
 Table~V of Mattila (\cite{mattila80a}) provides $UBV$ surface brightness 
estimates of the ISL, assumed here to be equal to the ISRF,
in the solar neighbourhood and in units of S10, where one S10 is one 
10th magnitude star per square degree. They
were converted into apparent $UBV$ magnitudes by integrating
over the sky (41253 sq. deg.). The corresponding fluxes $f(\lambda)$
derived with Allen's formulae are also given in Table~\ref{table3isrf}.

 Table~\ref{table3isrf} shows that these two ISRFs 
agree in $B$ and $V$ ISRF fluxes (and $B$-$V$ color),
but a difference of 0.45~mag. is found in the $U$ band (i.e. a
factor of 1.5), in the sense that the Mathis ISRF is  bluer in $U$-$B$.
 Mattila (\cite{mattila80a}, \cite{mattila80b})
  found  in his synthetized ISL a large flux jump 
between 3\,500~\AA \, and 4\,200~\AA \, which results from the Balmer 
jump of A-type stars and  the discontinuity at 4\,000~\AA \, 
from late-type stars. In contrast, the MMP83 ISRF is expressed by 
3 blackbodies $plus$ an UV component for $\lambda < $ 2\,500~\AA,
 and this approximation is said to be within 15 percent (or better)
 of the various observational data over the full spectral range 
(see MMP83, and also Mezger et al. \cite{mezger82}, their Appendix C). 
Therefore, at this stage, we find it interesting to keep both approaches
for further discussion.

\begin{table}
   \caption[]{Characteristics of the ISRF}
   \label{table3isrf}
   \begin{flushleft}
   \begin{tabular}{llll}
   \hline
   \hline
\noalign{\smallskip}
    $\lambda$   & $f(\lambda)$    &  magnitudes & Notes\\
    ($\mu$m) & (erg cm$^{-2}$ s$^{-1}$ \AA$^{-1}$) & & \\
\noalign{\smallskip}
     \hline
\noalign{\smallskip}
\multicolumn{4}{c}{ from Mathis, Metzger \& Panagia (\cite{mathis83})}\\
\noalign{\smallskip}
 0.346  & 1.30\,E-6   & $U$ = $-$6.21      & (1)\\
 0.435  & 1.50\,E-6   & $B$ = $-$5.89      & (1)\\
 0.550  & 1.57\,E-6   & $V$ = $-$6.54      & (1)\\
        &           & $U-B$ = $-$0.32  & \\
	&           & $B-V$ = +0.65  & \\
  \hline
\noalign{\smallskip} 
\multicolumn{4}{c}{ from Mattila (\cite{mattila80a})}\\
\noalign{\smallskip}
 0.365  & 0.87\,E-6   & $U$ = $-$5.76      & (2)\\
 0.440  & 1.47\,E-6   & $B$ = $-$5.87      & (2)\\
 0.550  & 1.52\,E-6   & $V$ = $-$6.52      & (2)\\
        &           & $U-B$ = +0.11  & \\    
	&           & $B-V$ = +0.65  & \\

    \hline 
\noalign{\smallskip}
   \end{tabular}\\
\end{flushleft}
Notes: 
(1) $UBV$ are derived with:\\
 $U=-2.5~ log$$_{10}$$f(0.365) -20.925$\\
 $B=-2.5~ log$$_{10}$$f(0.44) -20.450$\\
 $V=-2.5~ log$$_{10}$$f(0.55) -21.050$\\
(2) $f(\lambda)$ derived from $UBV$ data with above formulae 
\end{table}

\subsection{Parameters for the dust envelope of IRC+10216}

Concerning the CSE of IRC+10216, we have adopted as a baseline
the results of Groenewegen (\cite{groenewegen97},
 herafter G97), who considered a  large 
number of constraining observations (that probe mostly, but not
exclusively, the inner layers of the CSE).
The adopted parameters 
are listed  in Table~\ref{table4irc}. 
Although the CSE is known to show multiple shells 
(MH99, Mauron \& Huggins \cite{mauron00}),
 we  assume a smooth and spherically 
symmetric density distribution and will attempt to reproduce 
azimuthally-averaged profiles, leaving the consequences of the 
non-radial envelope  structure
for future study. With an assumed distance of 135~pc, G97 estimates a
dust mass-loss rate of $1.1~ 10^{-7}$\Msolyr beyond 4.47 stellar radii 
(inner envelope radius). Furthermore, we considered an outer envelope
radius of 210'' ($R_{\rm out}$ $=$ 6\,000~$R_*$) corresponding to 
the most distant material
detected by Mauron \& Huggins (\cite{mauron99}). This is certainly a 
lower limit for the 
IRC+10216 envelope outer radius since Young et al. (\cite{young93})
 found dust twice as 
far away using  IRAS data. However, we checked that our results concerning
the model brightnesses and colors of the IRC+10216 nebula are 
independent of  $R_{\rm out}$. Considering a larger outer radius would
only lead to a slightly larger width of the brightness profiles (see
the discussion below).

As recommended by G97, 
the grains considered here are 
 spherical particles with a single size ($a$ = 0.16 $\mu m$), and
their composition is that of  amorphous carbon with optical properties
given by Rouleau \& Martin (\cite{rouleau91},
 labelled AC1 in their Table~1; see also Appendix C). 
This table also  indicates another possible 
choice for the grain radius, $a$ = 0.05 $\mu m$ (as found by MR87 and 
discussed below).
It has to be noted that silicon carbide grains are also found in 
the envelope of IRC+10216. However, since (i) only a few percent of 
the dust grains are found
in SiC form, and (ii) their opacity effects are mostly seen around
11~$\mu$m and are completely negligible in the optical, we do not consider 
such  grains in our model. 

With the parameters of Table~\ref{table4irc} 
(standard model, with $a = 0.16 \mu m$), 
the radial optical thickness of the envelope in $B$ (taken from
the inner to the outer radius of the envelope)   
$\tau_{B,{\rm rad}}$ is equal to 84.3, in excellent agreement
with G97 if he neglected SiC grains. The observed
spectral energy distribution of IRC+10216 is well reproduced by our model
except for some small departures around 11~$\mu$m in the SiC signatures.

\begin{table}
   \caption[]{Characteristics of the envelope of IRC+10216}
   \label{table4irc}
   \begin{flushleft}
   \begin{tabular}{ll}
   \hline
   \hline
 \noalign{\smallskip}
     Parameter                      &  Value            \\
\noalign{\smallskip}
   \hline
 \noalign{\smallskip}  
\multicolumn{2}{c}{\it Standard model (0.16-$\mu m$ grains)}\\
Distance  $d$                   &  135 pc           \\
Central star $T_{\rm eff}$ & 2000~K \\
Central star radius $R_*$       & 35 mas            \\
Dust density law                & $\rho _{d} \propto r^{-2}$\\
Inner envelope radius $R_{\rm in}$   & 4.47 $R_*$ \\
Outer envelope radius $R_{\rm out}$   & 6\,000 $R_*$ \\
Dust mass loss  \Mdot$_{d}$     &  1.1\,10$^{-7}$ \Msolyr \\
Grain velocity  $v_d$           &  17.5 \kms        \\  
Grain radius    $a$             &  0.16 $\mu m$     \\  
Grain density  $\rho_d$         &  2.0 g cm$^{-3}$  \\ 
Dust optical properties           & amorphous carbon (1)\\
Radial opacities $\tau_U$, $\tau_B$, $\tau_V$ & 77.1, 84.3, 80.2 
\,\,\,\,\,\,\, (2)\\
\noalign{\smallskip}
\hline
\noalign{\smallskip}
\multicolumn{2}{c}{\it Small grain model (0.05-$\mu m$ grains)}\\
\multicolumn{2}{c}{\it Same parameters as above except:}\\
Grain radius $a$                 & 0.05 $\mu m$      \\
Radial opacities $\tau_{U}$,  $\tau_{B}$, $\tau_{V}$  & 66.7, 41.3, 24.0 
\,\,\,\,\,\,\, (2)\\ 
\noalign{\smallskip}
    \hline 
\noalign{\smallskip}
   \end{tabular}\\
\end{flushleft}
Notes: (1)  type  AC1 from Rouleau \& Martin (\cite{rouleau91});\\
(2) radial opacities like $\tau_{U}$ are also noted $\tau_{U,{\rm rad}}$ in 
the text
\end{table}

The optical thickness of the nebula {\em along the line of sight} at
$\theta$ arcseconds from the center
 can be written as:

\begin{equation}
\tau_{\lambda}(\theta) =  \pi \, \tau_{\lambda, \rm{rad}} \, \, 
 \frac{ \theta_{\rm in}}{\theta} \,\,\,  
q \, [ \frac {2}{\pi} \, arctan(\frac{\sqrt{1-u^2}}{u}) ] .
\end{equation}

Here, $\tau_{\lambda,\rm{rad}}$ is the radial optical thickness of the
envelope (i.e. from the center of the nebula up to $R_{\rm out}$);
$\theta_{\rm in} = R_{\rm in}/d $ corresponds to the angular 
inner radius of the envelope; $\theta_{\rm out} = R_{\rm out}/d $.
The factor $q$ is  very close to 1.0 and can be ignored: rigourously
 $q = 1/(1 - R_{\rm in}/R_{\rm out})$. The quantity
$u$ is $\theta / \theta_{\rm out}$, and the factor 
$ [ \frac {2}{\pi} \, arctan() ]$ is 
the effect of the outer envelope limit, and is equal to 1 if 
no outer limit is assumed. 

With the parameters of Table~\ref{table4irc}, and 
for an angular distance from the center of 50$''$, the standard  
model predicts  in the $B$ filter a line of sight opacity 
$\tau_B(50'') = 0.70$. The  factor in brackets due to the external
edge is 0.85 and an unlimited envelope with the same 
$\tau_{\lambda, {\rm rad}}$ would have an opacity 15 percent larger,
 i.e. 0.82 . At 100$''$, one
finds $\tau_B(100'') = 0.28$. Therefore, the  model envelope is  
optically thick in the observed region for offsets
 between 0 and $\sim$ 100-200 $''$, and a radiative transfer code 
is needed to predict the nebula optical
 surface brightness.

\subsection{The Monte-Carlo radiative transfer code}
In order to model ISRF illumination, we have developed a specific
radiative transfer code based on an upgraded version of the one
described in Lopez et al. (\cite{lopez95}).
Although the code has been built to allow 
for non-spherical dust density distributions, we consider in the present
 work spherically symmetric envelopes with 
a radial power-law  distribution of the form
 $N(r) = N_{in} \times (R_{in} / r)^{-n}$, 
 where $r$ is the radial distance from the
star with $R_{in} < r < R_{out}$, and  $N(r)$ is the number density
of grains.
The dust grains are assumed to be homogeneous and  spherical with a single 
radius $a$. Their optical properties, i.e. the extinction $Q_{\rm ext}$, 
scattering $Q_{\rm scat}$, and absorption $Q_{\rm abs}$ 
efficiencies, and the scattering phase function, 
are derived from optical indices using Mie theory. Finally,
the optical thickness of the envelope at the wavelength $\lambda$ is defined by
\begin{equation}
\tau_{\lambda, {\rm rad}} = \int_{R_{in}}^{R_{out}} N(r) \pi a^{2}\, 
Q_{\rm ext}(\lambda) \, dr
\end{equation}

Two sources of illuminating  radiation can be considered in the code:
 the first is the central star with radius 
$R_\star$  and assumed to radiate as a blackbody of a given effective 
temperature. This source is considered only for computing the spectral 
energy distribution (SED), found to be in full agreement with that of G97. 
The transfer of radiation in the dust and the radiative 
equilibrium of the grains are numerically solved by a Monte-Carlo method at
30 different wavelengths, and more details are given in Lopez et al. 
(\cite{lopez95}).\\

The second source is the ISRF, whose spectral energy distribution
is taken either from MMP83 or from Mattila (\cite{mattila80a}). 
The ISRF is  synthesized using  
a large  light-emitting  sphere centered on the nebula. 
The surface brightness of the sphere varies as cos$\phi$, where
$\phi$ is the inclination angle to its normal, 
and is equal for $\phi = 0$  to
 $f( \lambda ) / 4\pi$ (in erg cm$^2$ s$^{-1}$ \AA $^{-1}$ sr$^{-1}$),
 where $f( \lambda )$ is the ISRF value as 
given in Table~\ref{table3isrf}.
The radius of this ISRF sphere is  larger than the outer 
radius of the envelope, and it can be shown that
each point inside this ISRF sphere  receives in the absence of matter 
a homogeneous and isotropic photon flux equal to the ISRF.\\

In order to obtain a model image to compare with the $UBV$ observations
described above, we consider only ISRF photons emitted by the sphere
that have been actually scattered by dust grains in the 
envelope, and we ignore those directly emitted towards the observer and 
passing through the envelope without any interaction. 
This has to be done in order to avoid a bright background, which is not 
observed. The reason is that in the reality (but not in our model),
 the ISRF photons come mainly from 
individual point-like sources (the stars). The contribution of diffuse 
light scattered by the galactic cirrus and interstellar clouds is small, 
and this is  especially true for ISRF photons coming from behind IRC+10216,
 which is at $l=221 \degr$, $b = +45 \degr$. Within 
$\sim 2\degr$ of IRC+10216, little interstellar extinction is detected, 
since an  estimate $E(B-V) \leq 0.03$ mag can be obtained  from the maps 
of Burnstein \&  Heiles (\cite{burstein82})  and those of 
Schlegel et al. (\cite{schlegel98}). For 
such a low extinction, the observations of
Guhathakurta \& Tyson (\cite{guhathakurta89}) suggest that the diffuse
galactic background should be of the order of 27-28 B-mag arcsec$^{-2}$. This
is more than 25 times less than the ISRF surface  brightness when it is 
averaged over the sky (this spatial average is equal to 
23.44 $B$-mag arcsec$^{-2}$). Consequently, in order to
avoid in our model this fictitious, high background of 23.44 
$B$-mag arcsec$^{-2}$, 
direct photons, which are emitted by the ISRF sphere but are eventually 
not scattered by the envelope, are ignored.\\

 In the $UBV$ range considered here, 
the model envelope brightness is entirely due to scattered ISRF photons
since (i) its optical depth is so large that no  photons from the star can 
escape and (ii) the radiation emitted by the heated dust is insignificant
in visible light. Therefore, no central star was implemented in our simulation.
 For a given set of parameters and for each wavelength,
 the code returns an image of the externally illuminated envelope.
 Radial brightness profiles are  then derived.  To get a profile with a 
reasonably low noise, around 20 million events (``scattered photons'') 
are considered per 
Monte-Carlo simulation and the 
computation time in parallel mode is about 5 hours on 
a Compaq AlphaServer equipped with four 525\,MHz microprocessors.\\

\section {Model results, interpretation and discussion}
\label{SecInterpret}

\begin{figure}
\resizebox{\hsize}{!}{\rotatebox{-90}{\includegraphics{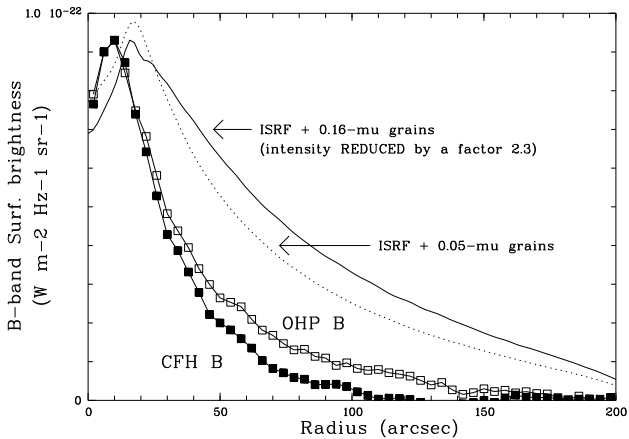}}}
\caption{Model brightness profile of IRC+10216 computed in $B$ 
(0.435~$\mu$m) (thin continuous line).The envelope  parameters  
are standard (Table~\ref{table4irc}, $ a = 0.16 $$\mu$m).
 For clarity of the plot, 
the intensity of this profile is drawn after arbitrarily dividing it by 
a factor 2.3 in order to fit the plateau intensity.
 The data from OHP and CFH  are 
also shown (dots), together with the actual model profile for $a = 0.05 \mu$m 
(dotted line).}
\label{fig3modelsb}
\end {figure}

We consider here the results of modelling when $standard$ envelope
parameters are adopted (i.e. Table~\ref{table4irc}, with 0.16 $\mu$m grains), and
compare them with the observations. We also discuss a model where
 the grain size  is changed to 0.05\,$\mu$m while  
keeping  the same dust mass loss rate and  other parameters.
This 0.05-$\mu$m model reproduces the observed SED reasonably well
(see Fig.~3 of G97), and its
radial opacity $\tau_B$ is equal to 41.3, which is obtained by scaling 
the standard opacity with appropriate $Q_{ext}$ and $a$ values 
($N(r) \propto a^{-3}$ and $\tau_{B} \propto  Q_{ext}/a$, see Eq.~2).\\

\subsection{Analysis of the central part of the nebula} 

The first issue concerns the intensity and shape of the nebula in its
central part, within $\sim$ 50$''$ of the center.
The $B$ band is particularly suitable for consideration because there is no 
difference  between Mathis's and Mattila's ISRFs in $B$, and because the 
 plateau  in $B$ is free from a central bright source like in $V$. 
Figure~\ref{fig3modelsb}   shows that the agreement is poor between the standard
model and observations. In order to reproduce the  $B$-band  plateau 
maximum surface brightness, one has to divide
the model intensity by a factor 
of 2.3\,: this factor is clearly too large to be attributed to 
photometric measurement errors on the  plateau  brightness.
Secondly, concerning the shape itself, the
model profile is found to be much broader that the observed one. 
While the model predicts a radius at half intensity, $R_{1/2}$
of 73$''$, the observation gives $R_{1/2} = 29$$''$$\pm$ 2$''$.
One notes in Fig.~\ref{fig3modelsb} that there is some disagreement 
between the $B$ profiles
obtained at OHP and CFH, due to experimental difficulties 
described in Section~2, but this concerns large offsets and essentially
does not affect the  $R_{1/2}$ value of 29$''$.
Figure~\ref{fig3modelsb} also shows the $B$ profile 
for 0.05-$\mu$m grains: in this case,
there is a nice agreement on the  plateau  level, but the model profile
is still too broad, with $R_{1/2} = 54$$''$.\\ 

\begin{figure}
\resizebox{\hsize}{!}{\rotatebox{-90}
{\includegraphics{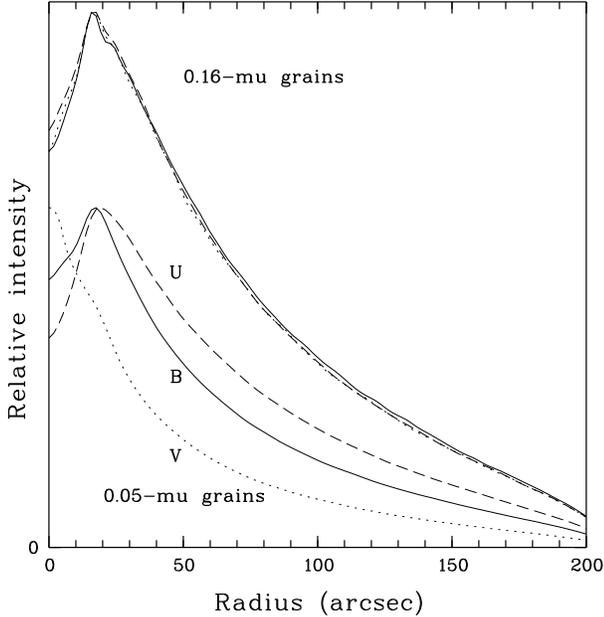}}}
\caption{Normalized model profiles for $UBV$ bands and for
grain sizes of 0.16 and 0.05 $\mu$m. The 0.16 $\mu$m profiles are
found to be almost similar (shape achromatism).
In contrast, the 0.05 $\mu$m profiles (shifted down by an arbitrary 
factor for clarity) clearly depend on wavelength.}
\label{fig4chromat}
\end {figure}

The second issue is the wavelength dependence of the nebula shape
(shape chromatism). Figure~\ref{fig4chromat} shows
the model profiles in $UBV$ for the two grain sizes. In each case, we have
normalized the $UBV$ profiles at the same (arbitrary) maximum value, so that
only the shapes are considered. In the standard case (0.16\,$\mu$m grains),
the shape is the same for $U$, $B$, or $V$  and the profiles
are exactly superposed. In contrast, if 0.05\,$\mu$m
grains are adopted, there is a strong chromatism in the sense that
the $U$ profile is much broader than the $V$ one. More precisely,
for a 0.16\,$\mu$m grain,
 $R_{1/2} = 73$$''$ for all filters, whereas for 0.05\,$\mu$m,
one has $R_{1/2} = 71$$''$, 54$''$, 29$''$ for $U,B,V$ respectively.
This situation is entirely due to the fact that small grains scatter 
nearly in the Rayleigh regime, with strong color effects, while larger 
grains are grey. Because the observations
show no chromatism (see Fig.~\ref{fig2profils})
 and negligible variation of $R_{1/2}$
with wavelength, the larger grain size (0.16\,$\mu$m) is clearly favoured.
It is  however true that for small grains  the $V$-band  $R_{1/2}$ 
value fits the observation (29$''$) rather well, but there is a 
large and increasing discrepancy for $B$ and $U$ bands.\\

\begin{figure}
\resizebox{\hsize}{!}{\rotatebox{-90}{\includegraphics{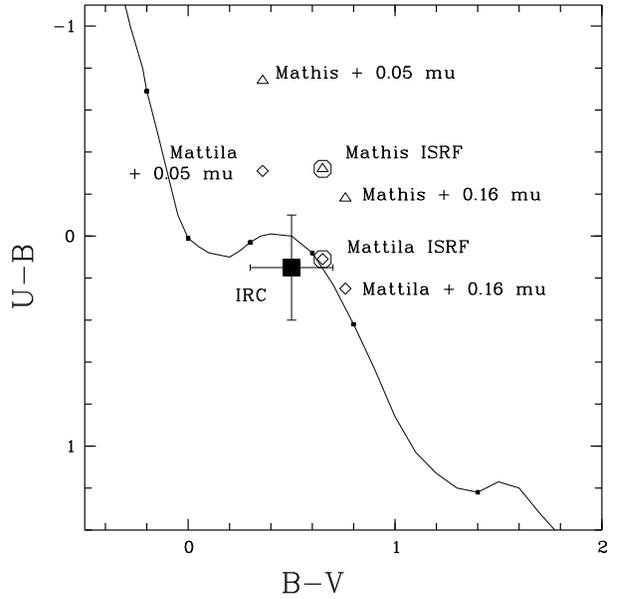}}}
\caption{
Comparison of the  plateau  $UBV$ colors  with models. The solid
black dot with error bars and labelled IRC
 represents the observed colors of the  plateau. The labelled triangles
 and lozenges represent the model plateau colors. Concerning the fields,
the ISRF of Mathis et al.  is represented by an encircled triangle, and
that of Mattila  by an encircled lozenge. 
 The line shows the $UBV$ locus 
of main sequence stars, with dots
indicating the positions of B3, A0, F0, G0, K0, and M0-type stars.}
\label{fig5colplateau}
\end {figure}

The third issue concerns the colors of the nebula.
Figure~\ref{fig5colplateau} shows that the observed  plateau  colors are relatively
 close to those  predicted for the Mattila's field scattered by  
0.16 $\mu$m grains.  Actually, the agreement is  better  between 
the observed colors and  the colors of Mattila's  field itself, 
suggesting  pure grey  scattering by 
circumstellar grains, i.e. large sizes. One can also see that 
if 0.05 $\mu$m 
grains are adopted, none of the fields match the 
observed colors. Adopting the field of Mathis and 
0.16 $\mu$m grains is  worse than adopting the 
field of Mattila and the same grains, especially when
considering the $U-B$ index for which the two fields differ.
We also note that  when one compares the colors of the fields 
with those of the main sequence  
stars, the Mathis field seems surprisingly blue in $U-B$, as blue as
an unreddened B8 star. If this was true, the rare stars bluer than B8
would have to compensate the many dwarf or giant stars redder than B8. 
It is much easier to understand the colors of  Mattila's field, 
which is located at an average position in $U-B$ and $B-V$.\\

\begin{figure}
\resizebox{\hsize}{!}{\rotatebox{-90}{\includegraphics{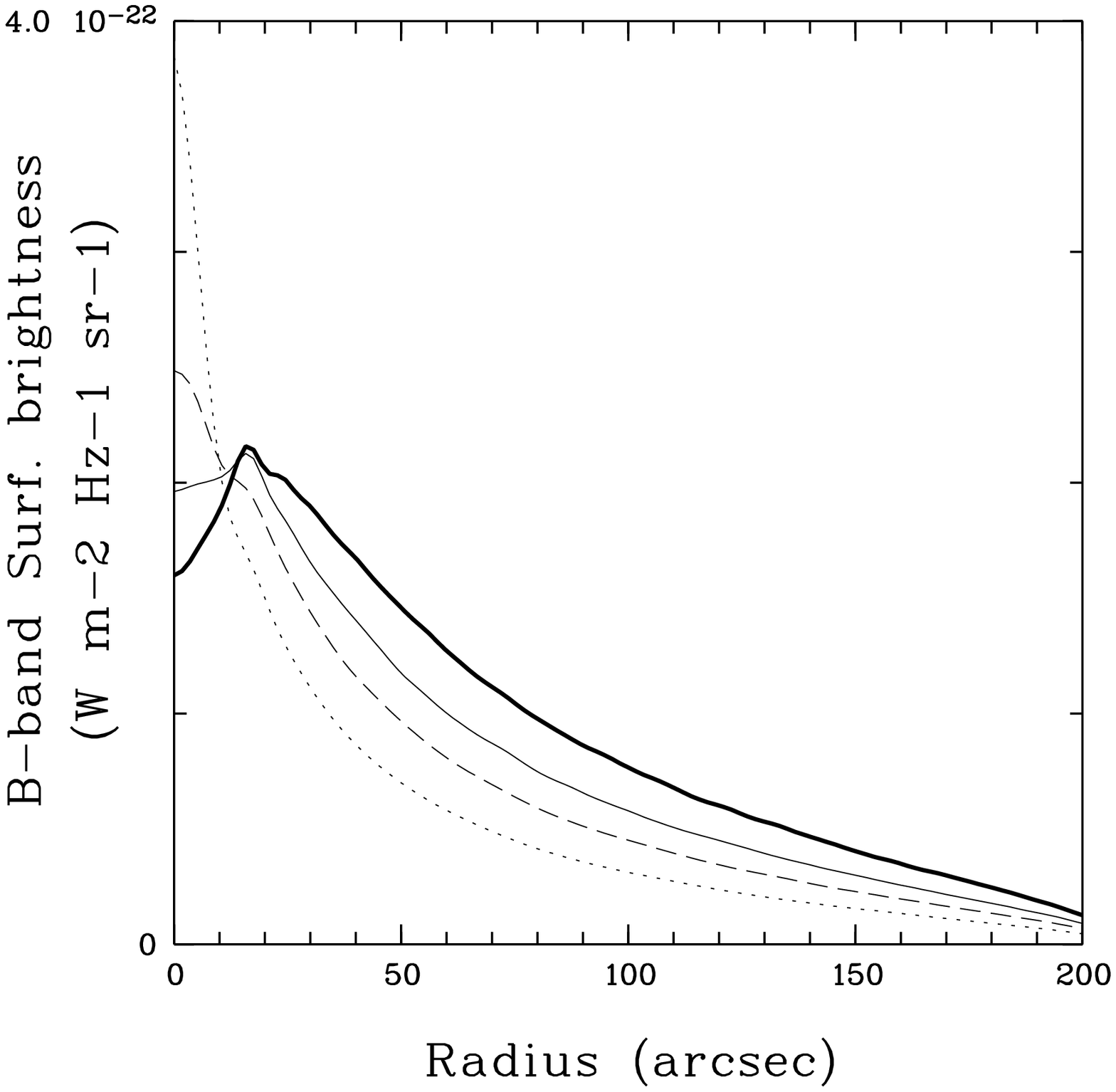}}}
\caption{ Evolution of the model profile in $B$ light 
 when $\tau_{B, {\rm rad}}$ is decreased from $\tau_{B, {\rm rad}}$=84
  (standard model,
 0.16$\mu$m grains, thick solid line),
to $\tau_{B, {\rm rad}}$=80 (thin solid line), $\tau_{B, {\rm rad}}$=60
 (dashed) and
$\tau_{B, {\rm rad}}$=45 (dotted).   Note that the $observed$  plateau  maximum 
brightness is at 0.92 10$^{-22}$\,W\,m$^{-2}$\,Hz$^{-1}$\,sr$^{-1}$}.
\label{fig6peak}
\end {figure}

Given the above considerations, it does not appear possible with our
data to settle the question of the dominant grain size. Both sizes
(0.05 and 0.16\,$\mu$m) present advantages and disavantages. The
small size fits reasonably well the plateau intensity and shape
{\it in the V band}, but fails for $B$ and $U$ (shape chromatism) and
for the plateau colors.  Regarding the large grain size (0.16 $\mu m$), the  
model intensity in $B$ is too strong by a factor of 2.3 on the plateau, 
and also too strong by a factor as large as $\sim 10$ for 
$\theta = 70$$''$ (Fig.~\ref{fig3modelsb}). It is interesting to note that
one cannot improve this situation by simply reducing the total opacity 
(or equivalently the dust mass loss rate) because this  produces an 
even stronger peak  at center and disappearance of the central depression, 
as shown in Fig.~\ref{fig6peak}. We have also explored grains larger than 0.16 $\mu$m: 
with AC1 material, a satisfactory value of $R_{\rm 1/2}$ 
is obtained for $ a =  0.28$ $\mu$m (then, in the $B$ band, 
$\tau_{B, {\rm rad}} = 34.4$)
 but the profile does not show any central depression,
and the plateau level is still two times too high compared to
observation.

 A better fit to the profile is obtained 
by  reducing progressively the dust mass loss rate of the standard model
 in the outer regions.
 For example,  a density decrease in $r^{-4}$ at $r > 20$$''$ produces 
a satisfactory $R_{1/2} = 29$$''$ (a $r^{-3}$ law is not enough), 
 although the plateau remains too high.

\subsection{Analysis of outer regions ($\theta \geq 70''$)}

There is another way of analysing the data:  instead of considering
the main part of the nebula and the plateau characteristics, 
one can focus on the outer layers, say at $\theta = 70$$''$.
At this radial distance, the surface brightness is certainly  less accurately
measured than at its center, but its faintness suggests that the envelope
can be considered as optically thin at this distance, which greatly 
simplifies the modelling.  Optical thinness is also supported by 
the fact that the 
density of background galaxies in a recent
 very deep $V$-band VLT image (de Laverny \cite{delaverny02})  is found 
 to be very similar at $\theta$ $\sim$ 70$''$ and  farther out. Then, a grain
located at this radius is illuminated by almost a pure ISRF for several 
reasons:  i.) the grain sees the central region of the nebula 
  over $\sim$ 50 degrees
FWHM, i.e. over a small solid angle of $\sim$\,0.6 sr, which is much 
less than 4$\pi$ sr,
and ii.) the average surface brightness of the nebula as seen by the grain
 is fainter than 25.7 mag/arcsec$^2$, which is  small against the average 
ISRF brightness of 23.44  mag/arcsec$^2$  in the $B$-band; iii.) most of 
the directions seen by the grain are optically thin. 
Consequently, in the optically thin regime and with an isotropic
illumination, the surface brightness of the dust envelope at a large
 angle $\theta$ can be written (see, e.g., Guhathakurta \& Tyson 1989):

\begin{equation}
 S_\lambda( \theta ) \, = \, \frac{1}{4 \pi} \, f_{\lambda}~~ \omega~~ 
 \tau_{\lambda}( \theta ) ,
\end{equation}

where $S_\lambda(\theta)$ has the units of ISRF flux  $f_{\lambda}$ 
per steradian,
$\omega$ is the grain albedo, and $\tau_\lambda(\theta)$ is the line of sight
optical opacity at offset $\theta$. For an unlimited $r^{-2}$ envelope, 
this opacity can be expressed as:

\begin{equation}
\tau_\lambda(\theta)  = \frac{3}{16}~~  \frac{Q_e}{a \rho_{\rm d}}~~
 \frac{\Mdot_{\rm d}}{v_{\rm d}~d}~~\frac{1}{\theta} ,
\end{equation}

and finally:

\begin{equation}
 S_\lambda( \theta ) = \frac{3}{64\pi}~ f_{\lambda}~ \frac{Q_{\rm sca}}{a}~ 
\frac{\Mdot_{\rm d}}{v_{\rm d}  \rho_{\rm d} d }~ \frac{1}{\theta} .
\end{equation}

In principle, the $\theta^{-1}$ dependance of Eq.~(5) might be
 checked on the observed profiles, but in practice, this is hardly
 possible because of the faintness of the signal and its uncertainties
 at these large offsets (see Sect.~2, and Figs.~\ref{fig2profils} \& 
 \ref{fig3modelsb}).
 
 In Eq.~(5), we can reasonably consider that the grain velocity $v_{\rm d}$, 
the grain density $\rho_{\rm d}$,
the distance $d$ and the ISRF strength   $f_{\lambda}$ are relatively well 
known (Table~\ref{table3isrf} and 4). Then, because the $B$-band observations provide 
at $\theta$=70$''$ a surface brightness  $S_B( \theta )$ of
$\sim$ 4.4 $10^{-20}$ erg cm$^{-2}$ s$^{-1}$ \AA$^{-1}$ arcsec$^{-2}$ 
(with an uncertainty of about a factor 2), one gets the following constraint:\\

\begin{equation}
\frac{\Mdot_{\rm d}}{1.1~~ 10^{-7}~ {\rm M_{\odot}~ yr^{-1}}} \times 
\frac{Q_{sca}}{a_{\mu}}  \approx 0.60
\end{equation}

For $a = 0.16$ $\mu m$ and $\Mdot_{\rm d} = 1.1~ 10^{-7}$  \Msolyr, which
is   the ``standard case'',
$Q_{sca}$ = 2.16 and $Q_{sca}/a_{\mu} = 13.5$. This  means
that the model flux is found to be $\sim$ 22 times larger than 
observed. However, in this case, the line of sight opacity
 $\tau_B(70'')$ is 0.58, so that the  optically thin approximation
is invalid here. Fig.~\ref{fig3modelsb} illustrates this case more rigourously 
and indicates a factor of $\approx$ 9 for the excess of model flux
at 70$''$.\\ 

In order to improve the agreement
between model and observations, a lower $\Mdot_{\rm d}$ or a
lower value for $Q_{sca}/a_{\mu}$ are needed, or both, i.e., less dust or dust
with a poorer scattering efficiency.
If  $ a = 0.05 \mu m$,  $Q_{sca}/a_{\mu}$ = 4.1 and Eq.(6) implies 
that $\Mdot_{\rm d}$ should be as low as 0.16~$10^{-7}$ \Msolyr, and then 
$\tau_B(70'')$ = 0.042. This ``small grain -- small dust loss" solution
can reproduce the faintness of the nebula at large offsets, but in no case
can it account for the central region when extrapolated down to
inner layers, because the radial opacity
of the whole envelope would be $\tau_{B, {\rm rad}} \approx 6$ 
($\tau_{B,{\rm rad}}$ is equal to 
$\frac{1}{\pi} \,\, \tau_B(\theta) \,\, \theta/\theta_{inn}$), 
largely insufficient to
produce a plateau. This solution cannot account also for the
observed polarized intensity in the near infrared reported by Tamura et
al. (\cite{tamura88}).

 Larger grains can also be envisaged, and for example
with $a = 0.28 \mu m$,  $Q_{sca}/a_{\mu}$ is also equal to 4.1; then
from Eq.(6), one finds again $\Mdot_{\rm d}$ $\sim$ 0.16~$10^{-7}$. 
In this case, $\tau_B(70'')$ = 0.035
and the envelope radial opacity is found to be
 $\tau_{B, {\rm rad}} \approx 5$, again 
largely insufficient to make a central plateau.

\subsection{Discussion}

The above analysis illustrates the fact that it is not easy to
reproduce with simple models all the envelope observations.
The standard model, which is based on the results of G97 and 
fits a lot of observations of dust in IRC+10216, especially in the 
infrared, fails to account for several features of the nebula seen 
in scattered galactic light. In particular, it seems difficult to 
simultaneously reproduce the plateau level, its profile, and  
the faintness of the  outer regions.\\

 These difficulties may be due, at least in part, to our simplified 
 treatment of dust. Grain  scattering  depends primarily on dust optical 
 properties (indices) and  the grain size. Scattering also depends 
 on factors like particle porosity and asphericity, and consequently 
 the applicability of the Mie theory. How large the effects of these last
 two factors are is difficult to estimate without extensive  calculations that 
 are beyond the scope of this work. We have also refrained from trying to
 search for a better agreement between model and observations by simply tuning 
 the optical properties, i.e., modifying at each wavelength the optical
 indices. The reasons are that  G97 used the Rouleau \& Martin
 AC1-type amorphous carbon indices, and, for consistency,
 the same  optical properties should be used here as far as possible;  
 secondly, the indices must obey the Kramers-Kronig relations and  cannot
 be arbitrarily modified in a limited wavelength domain. In summary, 
 our approach has been to consider only two different single particle
 sizes (0.16 and 0.05 $\mu m$), and these two cases correspond to 
 very different albedos and anisotropy $g$ factors: compared to the 
 standard case (0.16 $\mu m$ grains), the $0.05 \mu m$ grains
 have a lower albedo, which gives a very satisfactory fitting of 
 the plateau intensity, but these small grains produce too much chromatism 
 because their albedo strongly depends on wavelength (see above and 
 Appendix C for more details on adopted indices, albedos, etc). 
 It would certainly be interesting to envisage in future work a grain 
 size  distribution, although, to our knowledge, no robust observational
 constraint on circumstellar grain size distributions is presently 
 available (see also the arguments given by G97 for considering
 only a single grain size).\\

 Our analysis suggests that reducing the dust loss rate at  large offsets
 ($ > 25$ arcsec, corresponding to 1000 yrs ago) is a possible 
 solution to explain the nebula shape and
 its radial width. Variability of mass-loss with various time scales
 is a well-known characteristic of AGB stars (see for example
 the references in Olofsson \cite{olofsson99} or in Marengo et al. 
 \cite{marengo01}), and mass-loss
 variations for evolved carbon stars  is also theoretically predicted 
 (e.g. Wachter et al. \cite{wachter02}). Concerning 
 IRC+10216 itself, observations of its extended envelope in the CO 
 millimeter emission lines have been studied by Huggins et al.
 (\cite{huggins88}), and more recently by Groenewegen et al. 
 (\cite{groenewegen98}).  The last authors conclude, from their 
 analysis of the CO line profiles and maps, 
 that the  mass loss is enhanced by a factor of  $\sim$ 5 
 for $\theta > 50$$''$ compared to inner layers. 
 Such a density enhancement is clearly
 not seen in our radial profiles, and we checked that if one modifies 
 the standard model by increasing  $\Mdot_{\rm d}$ by $\sim$ 5 for 
 $\theta > 50$$''$, a huge bump appears in the envelope profile,
 and such a bump is definitively not observed (Fig.~\ref{fig2profils}
  \& \ref{fig3modelsb}). Thus,
 our data tend to suggest that the CO ``excess'' emission at large offsets 
 is either due, perhaps,  to incompletely  corrected instrumental lobe
 effects, or, if real, that they are the consequences of a physical
 phenomenon different from  a density enhancement,  
 as originally proposed  by Huggins et al. (\cite{huggins88}).\\

Concerning the faintness of the nebula, especially at large offsets,
it is also interesting to compare the present case of IRC+10216 
with the optical studies of high latitude clouds
(Guhatakurta \& Tyson \cite{guhathakurta89}). 
These clouds, that absorb and reflect 
the ISRF, are found to be, in the visible range, about 10 times fainter 
than predicted by models which reproduce their infrared radiation 
(with an {\it isotropic}
ISRF). The optical faintness of these clouds and that of
the outer layers of IRC+10216 might be related. Perhaps, if the ISRF is
not completely isotropic,  both phenomena are due to grain 
forward scattering of the photons coming preferentially from the galactic 
plane and escaping away from the observer's view.\\

As for the interstellar radiation field, it seems presently
premature to draw conclusions about its absolute strength 
(e.g. in the $B$ band) with some  accuracy, i.e. an accuracy 
better than a factor of 2, as was hoped originally when this work 
started.  MR87 proposed that, as a possibility, the field  
might be a factor $\sim$ 2.5 smaller than standard, but our  calculations 
with essentially the same envelope characteristics and the same grain 
radius (0.05 $\mu m$) do not confirm this point (Fig.~\ref{fig3modelsb}). 
We found no satisfactory reason to explain this inconsistency of results 
(see Appendix B for more details), but we do find that a fainter field 
improves the fit with 0.16 $\mu m$ or larger grains of AC1-type. 
Concerning  the field colors, our work suggests that the $U-B$
value of +0.11 found by  Mattila (\cite{mattila80a})
 is probably more correct than 
the one (-0.32) from  Mathis et al. (\cite{mathis83}).\\

Clearly, improving the simple model presented here appears desirable 
in order to reach (hopefully) firmer conclusions. One needs 
 especially to estimate 
the effects of a grain size distribution, and if possible of grain porosity
and asphericity.  The effect of a field anisotropy should be investigated.
 In addition, the influence of envelope homogeneities and 
structure in discrete shells will have also to be considered,
 since it is known that clumpiness  can  significantly  modify radiative 
 transfer in envelopes (e.g. Mathis et al. \cite{mathis02}).
In parallel with more sophistication in modelling, it would be fruitful 
to measure the colors of the nebula with less uncertainty, and extend 
the observations to red wavelengths (when the central variable star 
is at a minimum) and   space ultraviolet. 
Imaging other similar nebulae might also bring valuable information 
on the properties of amorphous carbon dust and on the ISRF
 in the Galaxy.\\

\section {Conclusions}

 We have presented new observations of the IRC+10216 nebula seen in the
ambient galactic light, including the first  $U$ imaging. 
Additional images in $B$ and $V$  were obtained as well. These data 
were used to establish a $UBV$ photometric sequence of nearby field stars, 
and to derive surface brightness profiles of the nebula, 
and $UBV$ colors of the  plateau. In blue light, where it is best observed,
the nebula has a central depression of 17\,\%, peaks at $\theta$ = 10$''$,
and its radius at half maximum $R_{1/2}$ is 29$''$. There is no sign of
an increased mass loss at $\sim$ 50$''$, as CO maps may indicate.\\

 The fact that the nebula shape, e.g. its  
half maximum intensity radius $R_{1/2}$,   does not depend on wavelength  
suggests that grains scatter the galactic light in the grey regime. 
This conclusion is  supported by the  plateau  colors 
which are  close to those of the ISRF as given 
by Mattila (\cite{mattila80a}).  The   grain size of  0.16\,$\mu$m 
recommended by Groenewegen (\cite{groenewegen97})
is consistent with these aspects of the data;
a smaller grain size like 0.05\,$\mu$m would produce  too strong a dependence 
of $R_{1/2}$ on wavelength and  too blue a color.
However, there remain important discrepancies between the 0.16$\mu$m grain
model and the observations, essentially concerning the intensity of the  
plateau  and the $R_{1/2}$ value, which are  better reproduced with 
0.05\,$\mu$m grains.\\

At first sight, our analysis suggests
that the mass loss may be lower in the outer nebula, but reaching firm 
conclusions is difficult before assessing the effects of a grain size 
distribution, the envelope structure and perhaps some ISRF anisotropy. 
Given these uncertainties on grain properties
and transfer modelling, there is also no firm evidence that 
the ISRF incident on 
IRC+10216  in the visible range is different from that 
usually taken in the solar neighbourhood.\\


\appendix

\section {$UBV$ photometry of field stars near  IRC+10216}
\label{appA} 
 In  Table~\ref{tablea1ubv}, $UBV$ photometry is given for 17 objects close to IRC+10216,
 derived from OHP observations. Coordinates (J2000) are from the APM catalog 
 (Irwin and colleagues; www.ast.cam.ac.uk/apmcat).
 Objects with flag 1, in  the last column,
 were used, when unsaturated, to calibrate the surface brightnesses
 of the IRC+10216 nebula in  images from OHP, CFH and VLT.
 Objects with flag~2 are much redder than those with flag~1 ($B-V > 1.2$), 
 and also redder than the  nebula; consequently, they were not used for 
 calibration. They are included here for completeness, as is 
 Object 12 with flag 3 which is an elliptical galaxy.

 Concerning the photometry, we do not pretend that its quality is the usual 
 one for standard photometric sequences (i.e. $\sim$ 0.01 mag), and our 
 initial goal was  more modestly to reach a $\sim$ 0.1 mag. uncertainty 
 on colors. The $UBV$ data  result from  aperture photometry made on
 only 2 to 5 images per color obtained during 3 nights (see the log in 
 Table~\ref{table1logobs}).
 Calibration on standard fields was limited in accuracy by the weather, 
 which was not quite photometric, except during the night of deep $U$ 
 1-hour exposures. The magnitudes and colors listed in  Table~\ref{tablea1ubv}
 were obtained by adopting plausible absorption coefficients for OHP, 
 by including a color term in the transformation from instrumental 
 to standard bandpasses, and by averaging the results from 
 different exposures. The listed uncertainty is the {\it internal} 
 error for colors, while it is 0.01-0.02 mag for $V$.

\begin{figure}
\resizebox{\hsize}{!}{\rotatebox{-90}{\includegraphics{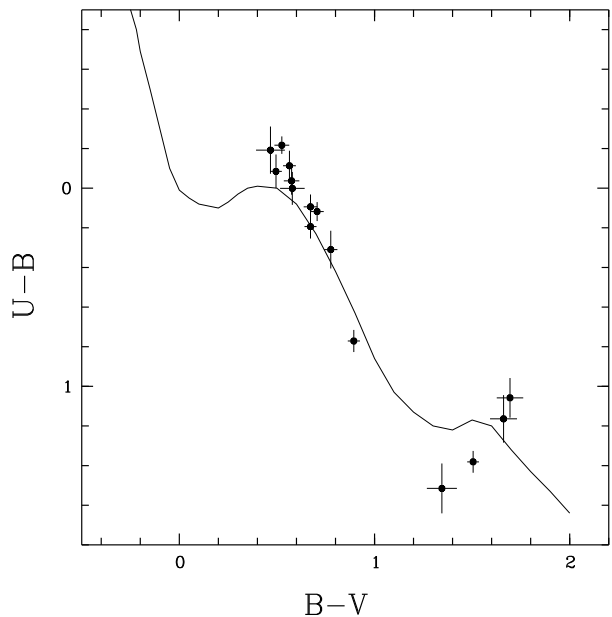}}}
\caption{ Color-color diagram for the field stars; the thin line is the
Population~I main sequence locus. Stars used for calibrating the nebula photometry
have $B$-$V$ $<$ 1.}
\label{figa1colcol}
\end {figure}

 A first independent verification of the OHP photometry in $V$ and $U$
 was achieved with the ESO VLT observations.
 The $V$ and $U-V$ values from VLT 
 were found to be within 0.04 and 0.03 mag of the OHP ones.
 A second verification was done
 by considering the color-color diagram of our photometry 
 (Fig.~\ref{figa1colcol}).
 This diagram shows  a group of
 10 stars with $B-V$ between 0.45 and 0.8 and located slightly 
($\sim$0.15 mag) above
 the location of the Population I main sequence (continuous line). 
This situation
 is fully consistent with expectations for  high latitude 
(IRC+10216 is at $b = +44 \degr$),
 faint ($V$=14 to 19) stars which statistically  probe 
 the metal-weak thick  disk.

 For instance, Ojha et al. (\cite{ojha99}) provide $U-B$ {\it vs} $B-V$ 
 diagrams and $B-V$
 histograms for  two fields at $b = +47 \degr$, and  
 another exemple of   
 $U-B$ {\it vs} $B-V$ diagram  for $b = +59 \degr$
  is given by Yamagata \& Yoshii (\cite{yamagata92}).
 The $B-V$ edge of the stellar locus from these works is at about 0.4 -- 0.45,
 exactly as in our data, and the $U-B$ average location is also in
 fair agreement with our work. Therefore, the final uncertainties 
 of this approximate 
 ``sequence" should be probably of $\pm$ 0.05 mag. for $V$,  and $\pm$0.05 to 
 0.1 mag for $B-V$ and $U-B$ respectively, at least for (not red) objects 
 flagged 1 used for calibrating the IRC+10216 surface brightness.
 The sequence  is not the main source of error in the nebula
 surface brightness photometry, whose accuracy (0.1 mag in $V$, 0.2-0.25 mag
 in color indices) is more limited by  its intrinsic faintness and  
 sky level subtraction.

\begin{table*}
   \caption[]{$U$$B$$V$ photometry of field stars near IRC+10216}
   \label{tablea1ubv}
   \begin{center}
   \begin{tabular}{rllllllll}
   \hline
   \hline
 Num   & $\alpha$(2000) & $\delta$(2000) &$V$ & $B$-$V$ & 
$U$-$B$ & $\delta_{\rm B-V}$&$\delta_{\rm U-B}$& flag \\    
   \hline
   
  1&  09 47 59.84 & $+$13  21  17.7 &   17.48 & $+$0.67 & $+$0.19 &    0.03  & 0.06&  1 \\
  2&  09 48 00.61 & $+$13  20  08.6 &   16.87 & $+$0.89 & $+$0.77 &    0.03  & 0.06&  1 \\
  3&  09 47 58.66 & $+$13  18  56.4 &   15.83 & $+$0.71 & $+$0.12 &    0.03  & 0.05&  1 \\
  4&  09 48 01.46 & $+$13  17  37.0 &   16.23 & $+$0.56 & $-$0.11 &    0.03  & 0.08&  1 \\
  5&  09 48 04.35 & $+$13  16  32.2 &   17.27 & $+$0.52 & $-$0.22 &    0.04  & 0.04&  1 \\
  6&  09 47 57.98 & $+$13  16  09.9 &   16.00 & $+$0.78 & $+$0.31 &    0.03  & 0.09&  1 \\
  7&  09 47 49.47 & $+$13  19  41.0 &   16.49 & $+$0.57 & $-$0.04 &    0.04  & 0.07&  1 \\
  8&  09 47 44.80 & $+$13  19  15.0 &   14.33 & $+$0.58 & $+$0.00 &    0.06  & 0.08&  1 \\
  9&  09 47 42.76 & $+$13  19  20.6 &   16.55 & $+$0.67 & $+$0.09 &    0.04  & 0.06&  1 \\
  10& 09 47 58.77 & $+$13  15  07.7 &   17.53 & $+$1.69 & $+$1.06 &    0.07  & 0.10&  2 \\
  11& 09 48 06.79 & $+$13  19  17.6 &   18.75 & $+$0.50 & $-$0.09 &    0.03  & 0.09&  1 \\
  12& 09 48 02.86 & $+$13  19  11.5 &   18.44 & $+$1.53 & $+$0.25 &    0.05  & 0.08&  3 \\
  13& 09 48 00.41 & $+$13  19  05.8 &   16.38 & $+$1.50 & $+$1.38 &    0.03  & 0.05&  2 \\
  14& 09 47 53.31 & $+$13  14  31.3 &   19.32 & $+$0.47 & $-$0.19 &    0.07  & 0.12&  1 \\
  15& 09 47 46.05 & $+$13  17  12.0 &   17.00 & $+$1.35 & $+$1.51 &    0.08  & 0.13&  2 \\
  16& 09 48 10.73 & $+$13  18  12.0 &   17.50 & $+$1.66 & $+$1.16 &    0.07  & 0.12&  2 \\
    
  \hline 
  \end{tabular}\\
~~~~\\
\end{center}
\end{table*}

%
%

\section{Quantitative verifications of the numerical code}

Some numerical tests have been done in order to verify the consistency
of the results returned by the code. They are in complete agreement
with other computations or analytical relations.

First, the computed SED of IRC+10216 with  the parameters of Table~\ref{table4irc}
is in excellent agreement with that published by G97. This increases confidence 
in the resolution of the radiative equilibrium equations.

Secondly, in order to check the computation of azimuthally-averaged 
brightness profiles (i.e. simulations of the ISRF illumination and scattering 
in the envelope), we considered the ideal case of an 
{\it optically thin} homogeneous sphere 
illuminated by an isotropic ISRF. We assume that the dust density is 
constant in the
sphere of external angular radius  $\theta_{\rm out}$. Because of
optical thinness,  each grain is illuminated by the ISRF itself, and it
can easily be shown that the surface brightness 
of such a sphere at an angle $\theta$ from its center can be expressed as:

\begin{equation}
S_{\lambda}(\theta) = \frac{1}{2 \pi} \, f(\lambda) \,
\frac{Q_{\rm sca}(\lambda)}{Q_{\rm ext}(\lambda)} \, \tau_{\lambda,{\rm rad}} \,
\sqrt{1-(\frac{\theta}{\theta_{\rm out}})^2}
\end{equation}
where $Q_{\rm sca}(\lambda)$ and $Q_{\rm ext}(\lambda)$ are  respectively
the scattering
and extinction efficiencies at wavelength $\lambda$, $f(\lambda)$ the 
ISRF flux and
$\tau_{\lambda, {\rm rad}}$ the radial optical thickness of the sphere
(i.e. from center to outer radius). The brightness  $S_{\lambda}(\theta)$
has the units of $f(\lambda)$ per steradian. An agreement
better than 1\% is obtained between the numerical and analytic brightness 
profiles for $\tau_{\lambda, {\rm rad}} \leq 0.01$, i.e. when the single 
scattering  assumption remains correct. This test validates 
our treatement of ISRF
illumination with an external sphere.

Thirdly, for an optically thin circumstellar envelope with a radial density 
varying as $r^{-2}$ , the brightness profile can also be analytically estimated
(for $\theta \ge \theta_{\rm inn}$):

\begin{equation}
S_{\lambda}(\theta) \, = \, \frac{1}{4} \, f(\lambda) \,
\frac{Q_{\rm sca}(\lambda)}{Q_{\rm ext}(\lambda)} \, \tau_{\lambda,{\rm rad}} \,
\frac{\theta_{\rm inn}}{\theta} \,  q$'$ 
\end{equation}

where $q' = 1$ if no outer limit is assumed:

\begin{equation}
 q' = (1-\frac{R_{\rm inn}}{R_{\rm out}})^{-1} 
\, \frac{2}{\pi} \, arctan(\frac{\sqrt{1-u^2}}{u})
\end{equation}

with $R_{\rm inn}$ and $R_{\rm out}$ being
the inner and outer envelope radius respectively, $\theta_{\rm out} = 
R_{\rm out} / d $
 and $u = \theta / \theta_{\rm out}$.
In this case also, the code returns very good results
in comparison to this analytical brightness profile, provided
that $\tau_{\lambda,{\rm rad}}$ is low enough.\\

We  have also examined
the case of similar but optically thicker spheres. For homogeneous spheres,
our simulations show that 
the brightness profiles depart from the above  equation (C.1) when 
multiple scattering plays a role, i.e. around $\tau_{\lambda, {\rm rad}} > 0.1$. 
An almost constant brightness profile with 
$\theta$ is obtained for $\tau_{\lambda,{\rm rad}} \sim 1$ and
bright-rimmed spheres with dark cores are found for optically thicker cases.
We also found that for optically thick spheres with radial density varying
as $r^{-2}$, the surface brightness shows a minimum at the center, reaches a 
maximum at some angular distance from the center and then decreases strongly.
All these results are entirely consistent with previous similar
simulations such as those found in Witt \& Stephens (\cite{witt74}, 
see their Fig.~1 \& 2).\\

Finally, we have attempted to exactly reproduce the 
numerical results of Martin \& Rogers (\cite{martin87}). 
One can first note that their Fig.~8 shows a significant 
chromatism of the plateau profile 
between the $I$ and the $V$ bands (0.81 $\mu$ and 0.55 $\mu$) that 
is qualitatively consistent with our findings explained in Sect.~5.
At 0.55\,$\mu m$, {\it for a standard ISRF}, their modelled plateau is at  
2.7 fu (flux unit, where 1 fu is 
$10^{-22}$ W m$^{-2}$ Hz$^{-1}$ sr$^{-1}$),  extends to
 $\theta \sim 5$$''$, and has $R_{1/2} \sim$ 25$''$.\\
 
Their model has essentially the same physical and
geometrical properties as ours, but unfortunately 
their optical grain properties are not given in an explicit way. It is
mentioned in their Sect. III-i that in $V$, $\omega$=0.38 and $g$=0.10.
In addition, from their Fig.~1 where $Q_{abs}/a$ and $Q_{sca}/a$ are plotted,
 one  can estimate $Q_{abs} \sim 0.21$ and $Q_{sca} \sim 0.15$. 
Using the Mie theory, with $a$=0.05$\mu$m and $\lambda$=0.55$\mu$m, 
one  finds corresponding indices  $n$=2.470  $k$=0.264, for which
$\omega$=0.384 and $g$=0.15, $Q_a$=0.221, $Q_s$=0.138, $Q_e$=0.359,
so that only $g$ is discrepant (we have found no reason for that
discrepancy). The model of Martin \& Rogers has
also $\tau_{11\mu}$=0.8 and from their fig.~1, $Q_e$(11$\mu$)=$Q_a$=0.0125.
Therefore, $\tau_{0.55\mu}$ = $\tau_{11\mu}$ $\times$
 $Q_e$($V$)/$Q_e$(11$\mu$) = 23.0,
 in fair agreement with the value of 21 mentioned by them.\\

With $\tau_{0.55\mu}$=23 and the $n$,$k$  values found above, our code
finds for a standard ISRF a very peaked central maximum at 2.75 fu, 
{\it with no plateau} and $R_{1/2}$ $\sim$ 15$''$.
Only  when $\tau_{0.55\mu}$ is increased to 30 does one get  a plateau
with maximum intensity of 2.0 fu, 
extended over 5$''$ and with $R_{1/2} \sim$ 36$''$. These two cases
 bracket the result of MR87, but are significantly
different. Therefore, there are some differences between the results of Martin 
\& Rogers and those obtained with our code, but it is difficult to say
whether they are due to differences in the codes themselves or due to
differences in indices.\\ 

\section{Grain optical properties}

Table~\ref{tablec1indices} lists the optical properties adopted in this work
(amorphous carbon of type AC1 from Rouleau \& Martin 
\cite{rouleau91}).\\

\begin{table*}
   \caption[]{Grain optical properties}
   \label{tablec1indices}
   \begin{center}
   \begin{tabular}{rlllllllll}
   \hline
   \hline
          &       &        &        & \multicolumn{3}{c}{$a = 0.16 \mu m$} 
                                    & \multicolumn{3}{c}{$a = 0.05 \mu m$}\\
\hline 
 Material & $\lambda$ ($\mu$m)&  $n$   &  $k$   & $Q_e$ &  $\omega$ & $g$ &  $Q_e$  &  $\omega$ & $g$\\    
   \hline
   
 AC1 &  0.365 (U) &  1.970 &  0.236 & 3.297 & 0.53 & 0.85 &  0.892 & 0.46 & 0.27\\
 AC1 &  0.435 (B) &  1.966 &  0.233 & 3.604 & 0.60 & 0.81 &  0.556 & 0.37 & 0.18\\
 AC1 &  0.550 (V) &  1.981 &  0.232 & 3.429 & 0.63 & 0.75 &  0.321 & 0.25 & 0.11\\
 \hline

  \end{tabular}\\
~~~~\\
\end{center}
\end{table*}

{\it Acknowledgments:}
The authors thank the anonymous referee and P.J. Huggins for comments
that helped us improve the paper. We also thank 
B. Gladman for careful reading of the manuscript.
Part of this work has been performed using the computing facilities
provided by the program ``Simulations Interactives et Visualisation en
Astronomie et M\'ecanique (SIVAM)'' at Observatoire de la C\^{o}te d'Azur. 
We  acknowledge support from the CNRS Program
{\it Physico-chimie du milieu interstellaire} \,(to N.M.), the
CNRS-INSU {\it Actions th\'ematiques innovantes} and the Minist\`ere de
l'Education Nationale et de la Recherche (to P.d.L.).


\end{document}